\documentclass[12pt,a4paper]{article}
\usepackage[T2A]{fontenc}
\usepackage[cp1251]{inputenc}
\usepackage[russian,english]{babel}
\usepackage{amsmath}
\usepackage{amssymb}
\usepackage{graphicx}
\usepackage{verbatim}
\usepackage{ifthen}
\usepackage{epsfig}
\usepackage{setspace}
\newcounter{fig}   \newcommand{\lbfig}[1]{\refstepcounter{fig}
\label{#1} }

\newcommand{\bea}{\begin{eqnarray}}
\newcommand{\eea}{\end{eqnarray}}
\newcommand{\be}{\begin{equation}}
\newcommand{\ee}{\end{equation}}

\def\bfph{{\pmb{\phi}}}

\newcommand{\re}[1]{(\ref{#1})}

\usepackage{graphicx}
\usepackage{graphpap,geometry}
\usepackage{epstopdf}

\title{Gauged Aloof Baby Skyrme Model}
\author{
{\large A.~Samoilenka}$^{\dagger}$
and {\large Ya. Shnir}$^{\dagger \star}$
 \\ \\
\\ $^{\dagger}${\small Department of Theoretical Physics and Astrophysics}\\
{\small Belarusian State University, Minsk 220004, Belarus}
\\ $^{\star}${\small BLTP, JINR, Dubna, Russia}
}

\begin{document}

\maketitle

\begin{abstract}
We present a study of $U(1)$ gauged modification of the 2+1 dimensional
planar Skyrme model with a particular choice of the symmetry breaking potential term which
combines a short-range repulsion and a long-range attraction. In the absence of the
gauge interaction the multi-solitons of the model are aloof, they consist of the individual constituents
which are well separated. Peculiar feature of the model is that there
are usually several different stable static multi-soliton solutions of rather
similar energy in a topological sector of given degree. We investigated the pattern of the solutions
and find new previously unknown local minima.
It is shown that coupling of the aloof planar multi-Skyrmions to the magnetic field strongly affects the pattern of
interaction between the constituents.
We analyse the dependency of the structure of the solutions, their energies and magnetic fluxes
on the strength of the gauge coupling. It is found that, generically, in the strong coupling limit
the coupling to the gauge field results in effective recovering of the rotational invariance of the configuration.
\end{abstract}

\section{Introduction}

The study of topological solitons in field theory can be traced back to
the seminal paper by Skyrme \cite{Skyrme:1961vq} where $SU(2)$-valued
non-linear model for atomic nuclei was suggested.
The Skyrme model can be derived from the expansion of the QCD
low energy effective Lagrangian in the large $N_c$ limit \cite{Witten},
then the topological charge of the multisoliton configuration is set into
correspondence to the physical baryon number.
Further, under certain assumption the semi-classical quantization of
rotations and iso-rotations of the Skyrmions allows us to get a good approximation to the
isospinning nuclei and describe the corresponding excitations which are associated with pions
\cite{Adkins,Brown:2009eh}. Thought, the Skyrme model has a limited success, there are several problems
with description of the nuclear masses since the interaction energy of the Skyrmions is much higher
than the corresponding experimental data for nuclei. Recent experimental observation of the heavy pentaquarks
\cite{Aaij:2015tga}, which could be considered as a sort of baryon-meson bound state
\cite{Karliner:2015ina} also cannot be explained in the conventional Skyrme model with the usual pion mass term.

Several modifications of the Skyrme model in 3+1 dimensions were suggested recently
\cite{Adam:2013wya,Sutcliffe:2011ig,Kopeliovich:2005vg,Gudnason:2015nxa}, mostly related with modification of the potential
of the model. Furthermore, the contribution of the Coulomb electromagnetic energy
is necessary to get a good agreement between the binding energies of heavy nuclei and the predictions of the
reduced BPS Skyrme model \cite{Adam:2013wya}. Therefore it is physically natural to extend the model
by gauging it to describe various electromagnetic processes of nucleons.

The $U(1)$ gauged Skyrme model was originally proposed in \cite{Callan:1983nx},
later the axially-symmetric gauged Skyrmions were considered in  \cite{Piette:1997ny,Radu:2005jp}.
It was noticed that
the gauging of a $U(1)$ subgroup may stabilize the
solitons even if the Skyrme term is dropped \cite{Schroers:1995he}, furthermore, in the gauged Skyrme model
the topological energy bound becomes saturated.

The planar reduction of the non-linear sigma model is known as
baby Skyrme model \cite{BB,Bsk}. This (2+1)-dimensional simplified model resembles the basic properties of the
genuine Skyrme model in many aspects. Furthermore,  the baby Skyrme model has a number of applications
on its own, e.g., in condensed matter physics where Skyrmion configurations were observed experimentally
\cite{CondMatt}, in the description of the
topological quantum Hall effect \cite{SkHall,Girvin}, or in brane cosmology where the solitons of the model
induce warped compactification of
the 2-dimensional extra space \cite{Kodama:2008xm}.
Also it was found that the restricted baby Skyrme model in 2+1 dimensions has
BPS soliton solutions saturating the topological bound \cite{Adam:2010jr}.

A peculiar feature of the planar Skyrme model is related with the
particular choice of the potential term which is necessary to
stabilize the solitons\footnote{ Note that a special choice of
the parameters of the baby Skyrme model allows to evade the
potential term and construct static soliton solutions in the
reduced model \cite{Ashcroft:2015jwa}, however dynamical
properties of these special configurations are still unknown.}. In
low-dimensional systems the effect of this term becomes more
significant than in the original Skyrme model, for example there
are different choices of the potential related with various ways
of symmetry breaking \cite{Ward,Hen,JSS,Salmi:2014hsa}. In
particular, a suitable choice for the potential term allows us to
separate the individual constituents of the planar Skyrmions, each
of them being associated with a fractional part of the topological
charge of the configuration \cite{JSS}. Another possibility is to
combine a short-range repulsion and a long-range attraction
between the solitons \cite{Salmi:2014hsa,Salmi:2015wvi}. Then the
multi-soliton configuration consists of aloof constituents.

Clearly, coupling the model to the electromagnetic field yields another possible channel of
interaction between the solitons.
Analysis of the gauged baby Skyrmions \cite{Gladikowski:1995sc} reveals very interesting features of the
corresponding solutions which  carry a non-quantized non-topological magnetic flux. Further, if the
Chern-Simons term is additionally
included in the Lagrangian, the planar Skyrmions become electrically charged \cite{Loginov}.
Recently, the properties of the soliton configurations in
the gauged BPS baby Skyrme model were investigated \cite{Adam:2014xfa}. An interesting observation is that
in the strong coupling limit the magnetic flux becomes quantized though there is no topological reasons for that
\cite{Gladikowski:1995sc,Shnir:2015twa}.

The aim of this paper is to discuss the structure of the soliton solutions of the
gauged aloof baby Skyrme-Maxwell system with
a rotationally invariant potential term analogous to that used in \cite{Salmi:2014hsa,Salmi:2015wvi}.
Our calculations are performed for multi-soliton solutions
up to charge 10.
We study numerically the dependence of masses of these configurations
and the corresponding magnetic fluxes on the gauge coupling constant,
both in perturbative limit and in the strong coupling limit without
any restrictions of symmetry.

\section{The model}
We consider a gauged version of the $O(3)$ $\sigma$-model with the Skyrme term in $2+1$ dimensions \cite{Bsk}
with a Lagrangian
\be
\label{lag}
{\cal L} = -\frac{1}{4}F_{\mu\nu}F^{\mu\nu}+\frac{1}{2}D_\mu \phi^a D^\mu \phi^a -
\frac{1}{4}(\varepsilon_{abc}\phi^a D_\mu \phi^b D_\nu \phi^c)^2 - V
\ee
Here $\phi^a$ denotes a triplet of scalar
fields, which is constrained to the surface of a sphere of unit radius: $\phi^a \phi^a=1$. We
introduced the usual Maxwell term with the field strength tensor defined as
$F_{\mu\nu}=\partial_\mu A_\nu -\partial_\nu A_\mu$. Clearly,
the scaling dimensions of this term and the Skyrme term, which is quartic in derivatives, are identical.

The coupling of the scalar field to the $U(1)$ gauge field $A_\mu$ is given by the covariant
derivative \cite{Schroers:1995he,Gladikowski:1995sc,Adam:2012pm}
\be
\label{cov-der}
D_\mu \phi^a= \partial_\mu \phi^a + g A_\mu \varepsilon_{abc} \phi^b \phi^c_\infty \, ,
\ee
where $g$ is the gauge coupling constant.
The localized field configuration has finite energy if $D_\mu \phi^a \to 0$, $F_{\mu\nu} \to 0$
and $V \to 0$ as $r \to \infty$.

Topological restriction on the field $\phi^a$ is that it approaches its
vacuum value at spacial boundary, i.e. $\phi^a_\infty = (0,0,1)$. This allows a one-point compactification of
the domain space $\mathbb{R}^2$ to $\mathbb{S}^2$ and
the field of the finite energy solutions of the
model is a map $\bfph:\mathbb{R}^2 \to \mathbb{S}^2$ which belongs
to an equivalence class characterized by the topological charge $Q = \pi_2(\mathbb{S}^2) = \mathbb{Z}$.
Explicitly,
\be \label{charge}
Q= - \frac{1}{4\pi}\varepsilon_{abc}\int \phi^a  \partial_x \phi^b \partial_y \phi^c ~dx dy
\ee

A special feature of the solutions of the planar Skyrme model is that their structure strongly depends on the
particular form of the potential term $V$. The most common choice is so called "old potential" \cite{Bsk}
\be \label{pot-old}
V = \mu^2 [1-\phi_3]\, ,
\ee
which is an analogue of the standard pion mass term in (3+1)-dimensional Skyrme model.
The symmetry is broken via the potential to $SO(2)$ and there is a unique vacuum
$\bfph_\infty = (0,0,1)$.
The corresponding solitons of degree
$Q=1,2$ are axially symmetric \cite{Bsk} however the rotational symmetry of the configurations of higher degree
becomes broken \cite{PZS}.

In the model with
double vacuum potential (or "easy-axis" potential) \cite{BB,Weidig:1998ii}
\be \label{double}
V = \mu^2 (1-\phi_3^2)\, ,
\ee
the multi-soliton solutions are rotationally invariant over entire range of values  of the mass parameter $\mu$.
The most general case of the one-parametric potential
\be \label{Karliner-Hen}
V = \mu^2(1 - \phi_3)^s
\ee
with $0 < s \leqslant 4$ was considered in \cite{Hen}.
Since the "old" potential \re{pot-old} corresponds to the attractive
force acting between the solitons, while the "holomorphic" potential $V = \mu^2(1 - \phi_3)^4$ is
repulsive \cite{Leese,Sutcliff-holo}, the parameter $s$ in the potential \re{Karliner-Hen} is
responsible for the balance of the
repulsive and attractive interaction between the Skyrmions.
Further, one can consider the linear combination of the
"old" and "holomorphic" potentials \cite{Salmi:2014hsa,Salmi:2015wvi}
\be \label{Salmi}
V = \mu^2\left[\lambda(1 - \phi_3) + (1-\lambda)(1 - \phi_3)^4\right] \, , \quad \lambda \in [0,1] \, ,
\ee
which corresponds to a short-range repulsion and a long-range attraction between the solitons.
Following \cite{Salmi:2014hsa,Salmi:2015wvi}, we restrict our consideration to the case $\lambda=0.5$.

The resulting
multi-soliton configuration is no longer rotationally invariant, each constituent of unit charge
is clearly separated from other, they form a cluster structure. Further, in the absence of the gauge interaction
there are usually several different static multi-soliton solutions of rather
similar energy in a topological sector of given degree  \cite{Salmi:2014hsa}. Note that this feature is in common
with Hopfion solutions of the Faddeev-Skyrme model \cite{Faddeev}.
Indeed, the structure of both models looks similar,
the corresponding Lagrangian \re{lag} includes the usual sigma model term, the Skyrme term, which is quartic in
derivatives of the field, and the potential term. Further,
the field of the Faddeev-Skyrme model in 3 spacial dimensions
is also a three-component unit vector restricted to the unit sphere $S^2$.
However, the domain space of the latter model is the compactified three dimensional
sphere $S^3$ and the Hopfion solutions are classified by the linking number associated with the
homotopy group $\pi_3(S^2) = \mathbb{Z}$. Thus, in some sense the baby Skyrme model can be considered as a planar
reduction of the Faddeev-Skyrme model \cite{Kobayashi:2013aza}.

In 2+1 dimensions the electric field of configuration is vanishing everywhere \cite{Gladikowski:1995sc}
and we can consider purely magnetic field generated by the Maxwell potential
\be
\label{el-ansatz}
A_0 = A_y = 0;\qquad A_x=A(x,y)\, ;
\ee
where  the gauge fixing condition is
used to exclude the $A_y$ component of the vector-potential. Thus the magnetic field is orthogonal to the $x-y$ plane:
$B = B_z = -\partial_y A_x$.
Note we do not use here the rotationally-invariant parametrization of the fields where the gauge fixing condition $A_r=0$ is
imposed \cite{Bsk,BB,Weidig:1998ii}.

The complete set of the field equations, which follows from the variation of the action of the baby Skyrme-Maxwell
model \re{lag}, can be solved when we impose
the boundary conditions.  As usually, they follow from the regularity on the boundaries
and symmetry requirements as well as
the condition of finiteness of the energy and the topology. In particular we have to take into account
that the magnetic field
is vanishing on the spacial asymptotic.  Explicitly, on the spacial boundary we impose
\begin{equation}
\label{bc}
\phi_{1}\biggl.\biggr|_{r \rightarrow \infty }\!\!\!\rightarrow
0\,,~~\phi_{2}\biggl.\biggr|_{r \rightarrow \infty
}\!\!\!\rightarrow 0\,,~~\phi_{3}\biggl.\biggr|_{r \rightarrow
\infty }\!\!\!\rightarrow 1\,,~~\partial_y A\biggl.\biggr|_{r \rightarrow \infty
}\!\!\!\rightarrow 0\, .
\end{equation}%
Here $r=\sqrt{x^2+y^2}$ is the usual radial variable.

As initial guess for further computation we used various combinations of single solitons with preassigned
phases and positions. These field configurations can be constructed via usual parametrization of the unit
scalar triplet $\phi^a$ in terms of the complex field $W$ via the stereographic projection
\be
\label{input}
W=\frac{\phi_1+i\phi_2}{1-\phi_3} \, .
\ee
Let $W_i(z)$, $z=x+iy$,  be the field of a single Skyrmion of degree one with an
arbitrary position and phase, then
the function  $\frac{1}{W}=\sum\frac{1}{W_i}$, $i=1,2,..Q$ allows us to obtain the field of
multi-soliton configuration of degree $Q$ with various relative phases and separations. For example,
the function
\be
W(z) = \frac{4}{(z-2)^{-1} -(z-1)^{-1}+ z^{-1} - (z+1)^{-1}+(z+2)^{-1}}
\ee
yields the input configuration in the sector of degree $Q=5$, which represents the chain of baby Skyrmions
with opposite relative orientations and separation between the constituents $d=1$.

\section{Numerical results}
In this section we outline our method for calculating the multi-soliton solutions of the gauged baby Skyrme model.
The numerical calculations are mainly performed on a equidistant square grid,
typically containing $160^2$ lattice points and with a lattice spacing $dx=0.15$
To check our results for consistency we also considered the lattice spacings $dx=0.1,0.2$.

\begin{figure}[hbt]
  \begin{center}
\includegraphics[height=4.cm,angle=00,bb=0 0 357 329]{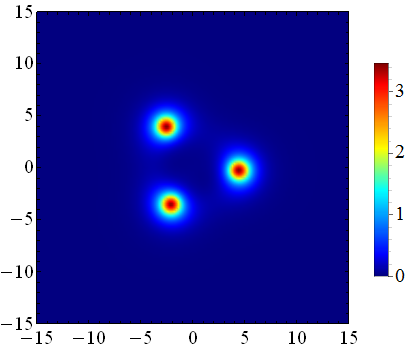}
\hspace{0.2cm}
\includegraphics[height=4.cm,angle=00,bb=0 0 357 329]{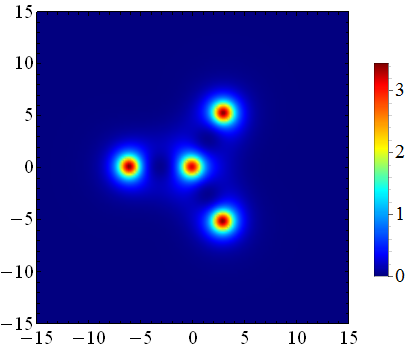}
\hspace{0.2cm}
\includegraphics[height=4.cm,angle=00,bb=0 0 357 329]{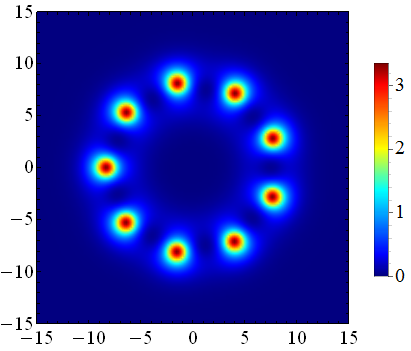}
\end{center}
\caption{Contour energy density plots of the $3 D_3$, $4 D_3$ and
$9 D_9$ of planar Skyrmions in the model \re{lag} at $g=0$.
}
\lbfig{fig:1}
\end{figure}

To construct multi-soliton solutions of the model \re{lag} we minimize the corresponding
rescaled energy functional of the static  configuration
\be
\label{eng-tot}
E= \int\left\{\frac12 B^2 +
\frac12 D_i \phi^a \cdot D_i\phi^a +
\frac{1}{4}(\varepsilon_{abc}\phi^a D_i \phi^b D_j \phi^c)^2
 + V\right\}~dx dy
\ee
with the aloof baby Skyrmions potential \re{Salmi}.
The gauge coupling $g$ is a parameter, each of our simulations began
at $g=0$ at fixed value of $\mu$ and $\lambda$,
then we proceed by making small increments in $g$.
For comparative consistency with \cite{Salmi:2014hsa} in most of our calculations
we choose $\mu^2=0.1$ and $\lambda=0.5$.

The numerical algorithm employed was similar to that used in \cite{Hale:2000fk}.
Well-chosen initial configurations of given degree were evolved
using the Metropolis method to minimize the energy functional \re{eng-tot}.
In our numerical calculations we
introduce an additional Lagrange multiplier to constrain the field to the surface of unit
sphere.
Simulations were considered to have converged to local minima if the quantity $-\frac{1}{E}\frac{dE}{dt}$,
where $t$ is the time of computation in minutes,
was less than $10^{-3}$. We also verify that the evaluated topological charge of the configuration is in
agreement with the input integer value.

\begin{figure*}[h!]
\begin{center}
\begin{tabular}{|c|c|c|c|c|}
\hline
Charge & \multicolumn{4}{c|}{g}  \\
\hline
1 & 0 & 0.4 & 1 & 2\\
& \includegraphics[height=2.4cm,angle=00,bb=0 0 357 329]{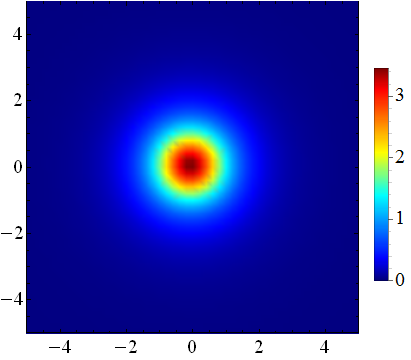} & \includegraphics[height=2.4cm,angle=00,bb=0 0 357 329]{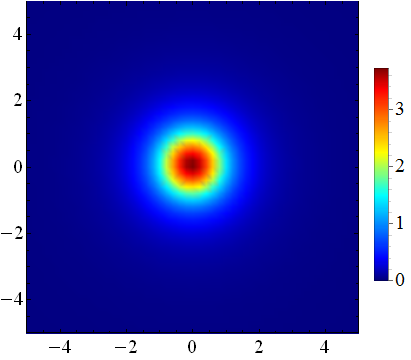} & \includegraphics[height=2.4cm,angle=00,bb=0 0 357 329]{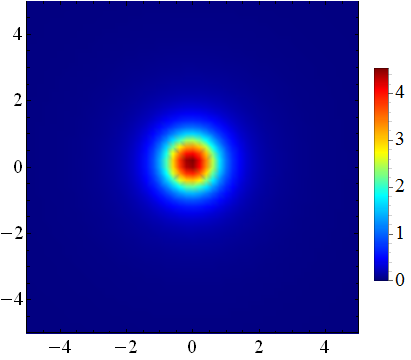} & \includegraphics[height=2.4cm,angle=00,bb=0 0 357 329]{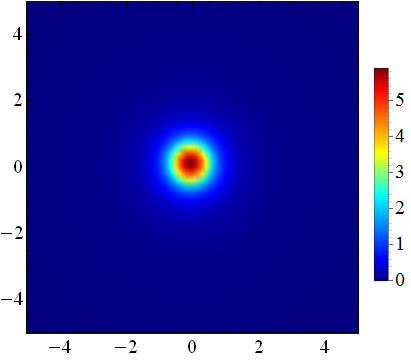} \\
\hline
2 & 0 & 0.15 & 0.3 & 1\\
&  \includegraphics[height=2.4cm,angle=00,bb=0 0 357 329]{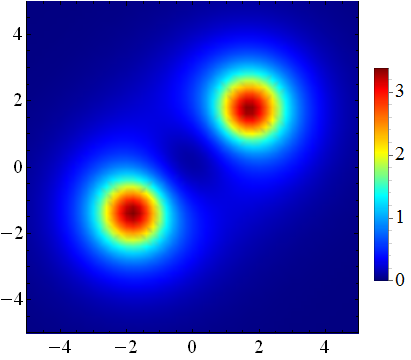} &  \includegraphics[height=2.4cm,angle=00,bb=0 0 357 329]{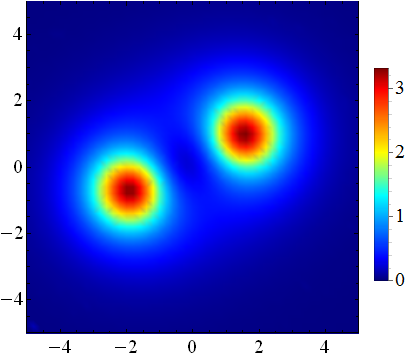} &  \includegraphics[height=2.4cm,angle=00,bb=0 0 357 329]{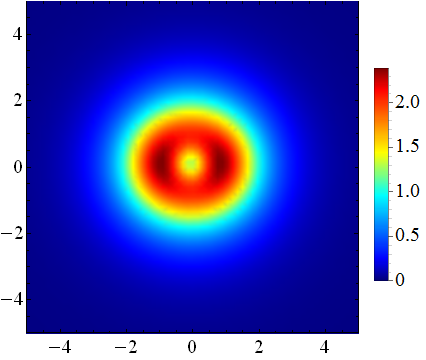} & \includegraphics[height=2.4cm,angle=00,bb=0 0 357 329]{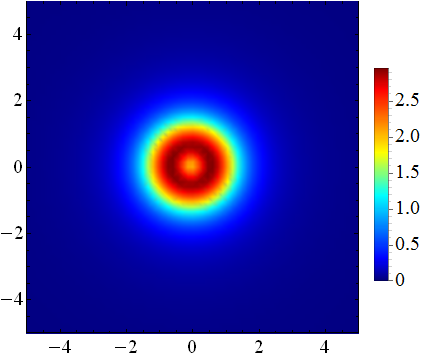} \\
\hline
3 & 0 & 0.3 & 0.6 & 2\\
& \includegraphics[height=2.4cm,angle=00,bb=0 0 357 329]{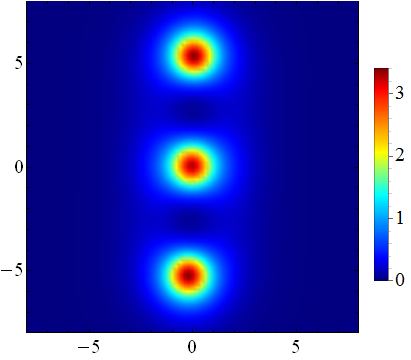} & \includegraphics[height=2.4cm,angle=00,bb=0 0 357 329]{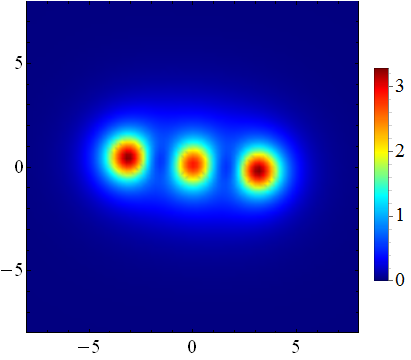} & \includegraphics[height=2.4cm,angle=00,bb=0 0 357 329]{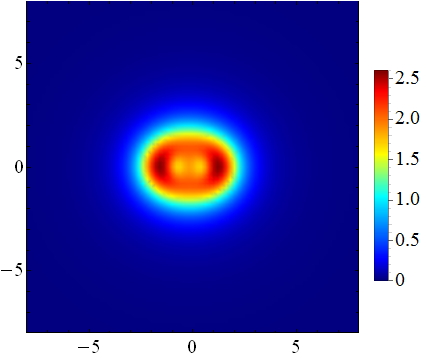} & \includegraphics[height=2.4cm,angle=00,bb=0 0 357 329]{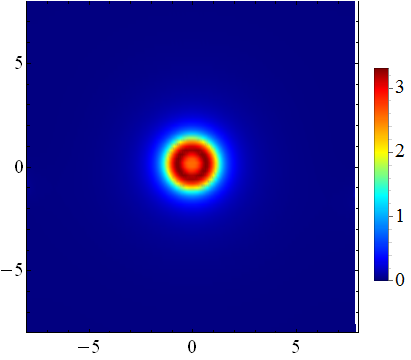} \\
\hline
4 & 0 & 0.4 & 0.6 & 1\\
&\includegraphics[height=2.4cm,angle=00,bb=0 0 357 329]{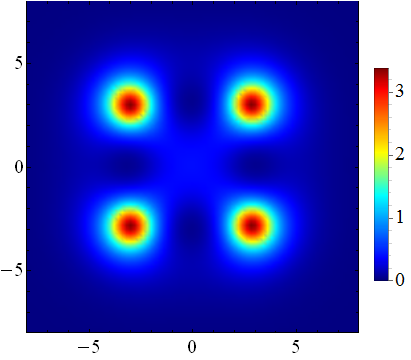} & \includegraphics[height=2.4cm,angle=00,bb=0 0 357 329]{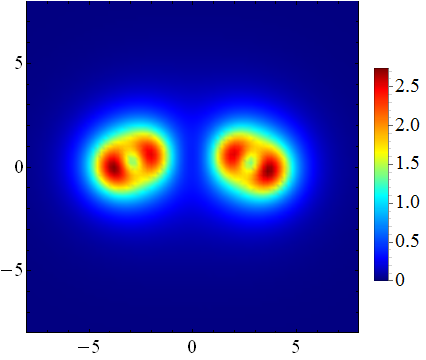} &\includegraphics[height=2.4cm,angle=00,bb=0 0 357 329]{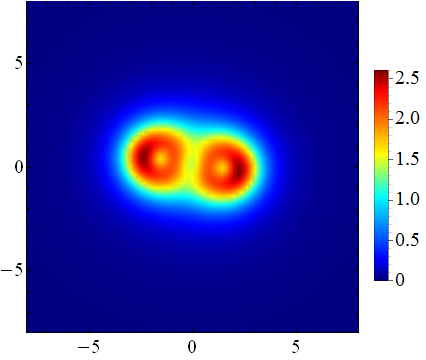} & \includegraphics[height=2.4cm,angle=00,bb=0 0 357 329]{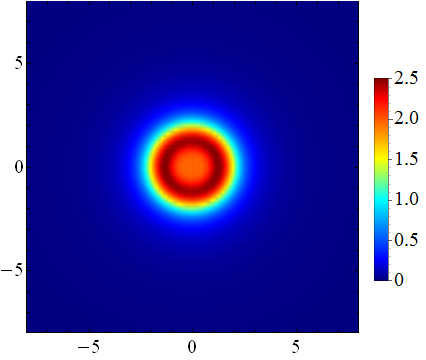} \\
\hline
5 & 0 & 0.4 & 0.6 & 1\\
&\includegraphics[height=2.4cm,angle=00,bb=0 0 357 329]{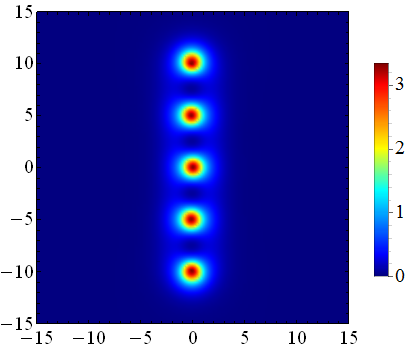} & \includegraphics[height=2.4cm,angle=00,bb=0 0 357 329]{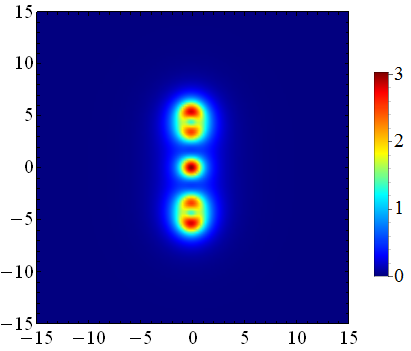} & \includegraphics[height=2.4cm,angle=00,bb=0 0 357 329]{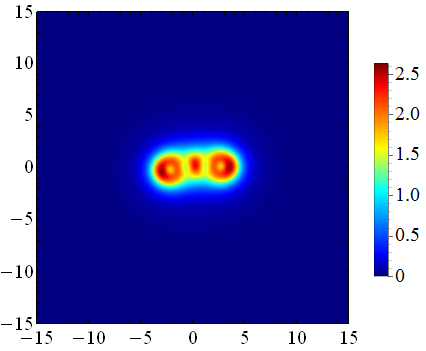} & \includegraphics[height=2.4cm,angle=00,bb=0 0 357 329]{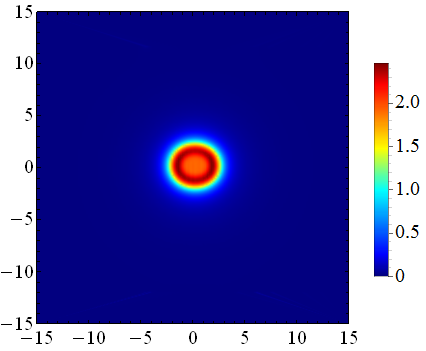} \\
\hline
6 & 0 (global) & 0.4 & 0.6 & 2 \\
& \includegraphics[height=2.4cm,angle=00,bb=0 0 357 329]{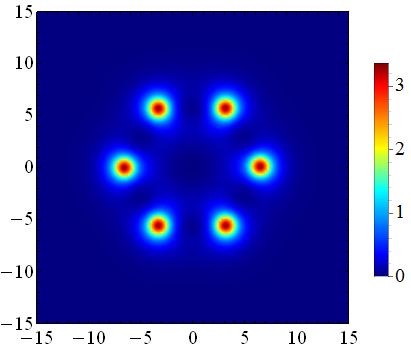} & \includegraphics[height=2.4cm,angle=00,bb=0 0 357 329]{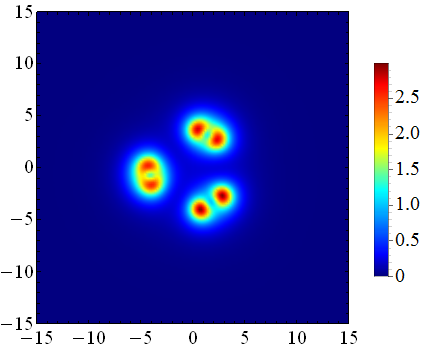} & \includegraphics[height=2.4cm,angle=00,bb=0 0 357 329]{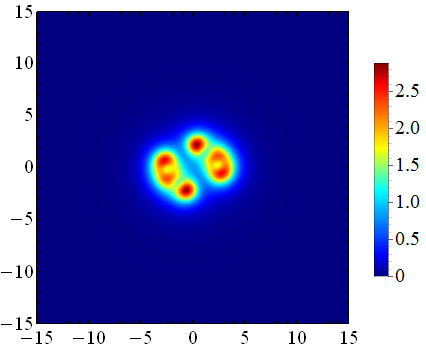} & \includegraphics[height=2.4cm,angle=00,bb=0 0 357 329]{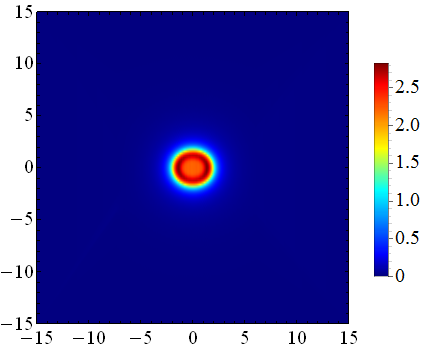} \\
\hline
6 & 0 (local) & 0.4 & 0.6 & 2 \\
& \includegraphics[height=2.4cm,angle=00,bb=0 0 357 329]{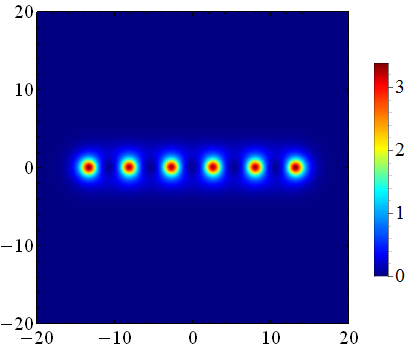} & \includegraphics[height=2.4cm,angle=00,bb=0 0 357 329]{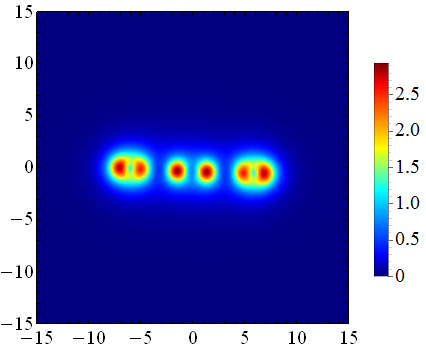} & \includegraphics[height=2.4cm,angle=00,bb=0 0 357 329]{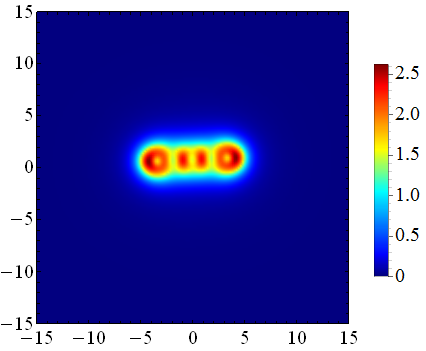} & \includegraphics[height=2.4cm,angle=00,bb=0 0 357 329]{ch6g2o0.png} \\
        \hline
\end{tabular}
\end{center}
\end{figure*}
\begin{figure*}
\begin{center}
\begin{tabular}{|c|c|c|c|c|}
        \hline
            Charge & \multicolumn{4}{c|}{g}  \\
        \hline
            7 & 0 (global) & 0.4 & 0.6 & 1 \\
& \includegraphics[height=2.4cm,angle=00,bb=0 0 357 329]{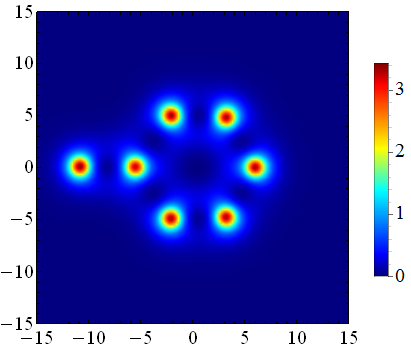} & \includegraphics[height=2.4cm,angle=00,bb=0 0 357 329]{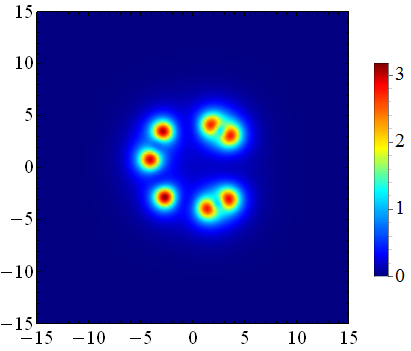} & \includegraphics[height=2.4cm,angle=00,bb=0 0 357 329]{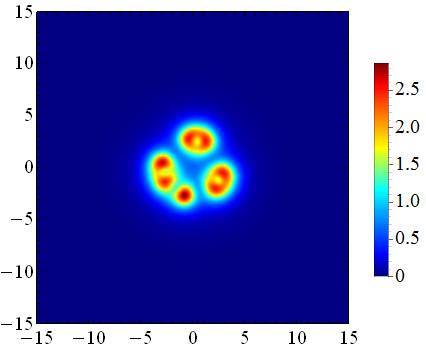} & \includegraphics[height=2.4cm,angle=00,bb=0 0 357 329]{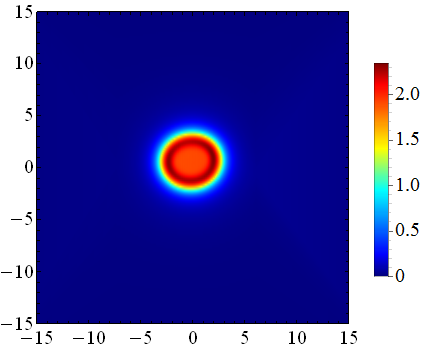} \\
        \hline
            7 & 0 (local) & 0.4 & 0.6 & 1 \\
& \includegraphics[height=2.4cm,angle=00,bb=0 0 357 329]{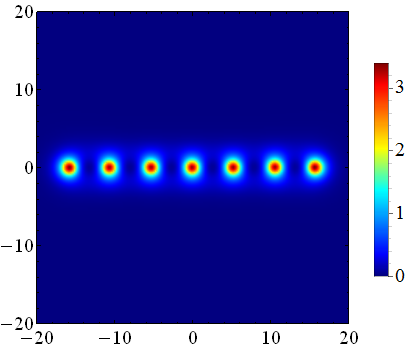} & \includegraphics[height=2.4cm,angle=00,bb=0 0 357 329]{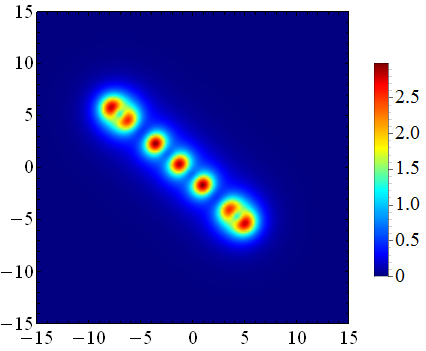} &\includegraphics[height=2.4cm,angle=00,bb=0 0 357 329]{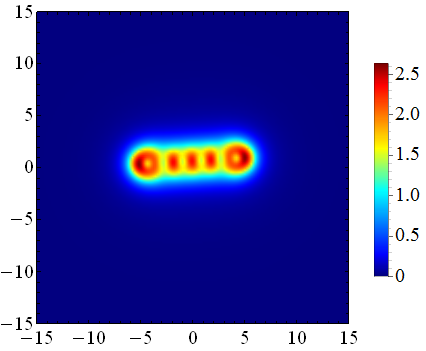} & \includegraphics[height=2.4cm,angle=00,bb=0 0 357 329]{ch7g1o0.png} \\
        \hline
            8 & 0 & 0.4 & 0.6 & 1\\
& \includegraphics[height=2.4cm,angle=00,bb=0 0 357 329]{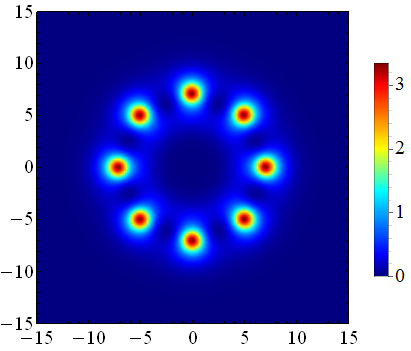} & \includegraphics[height=2.4cm,angle=00,bb=0 0 357 329]{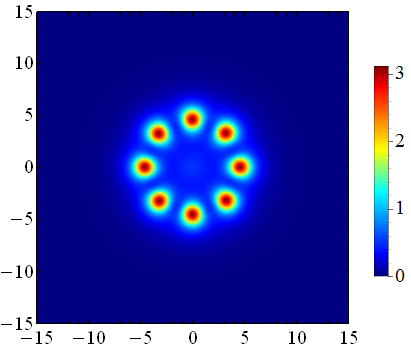} & \includegraphics[height=2.4cm,angle=00,bb=0 0 357 329]{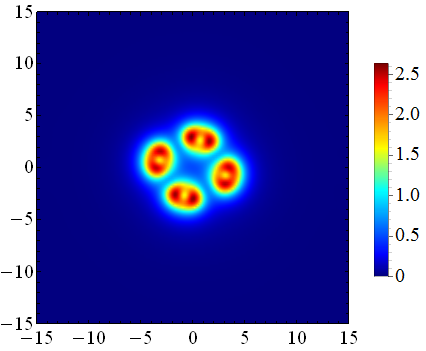} & \includegraphics[height=2.4cm,angle=00,bb=0 0 357 329]{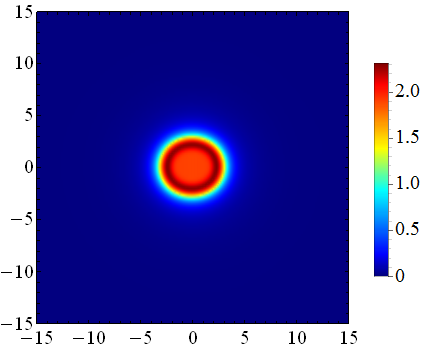} \\
        \hline
            8 & 0 & 0.4 & 0.6 & 1\\
& \includegraphics[height=2.4cm,angle=00,bb=0 0 357 329]{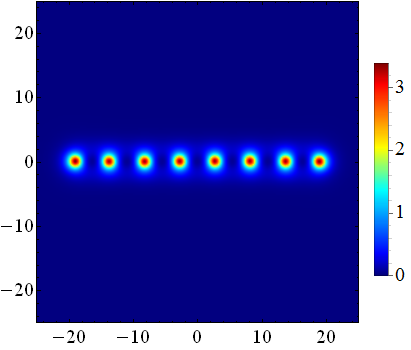} & \includegraphics[height=2.4cm,angle=00,bb=0 0 357 329]{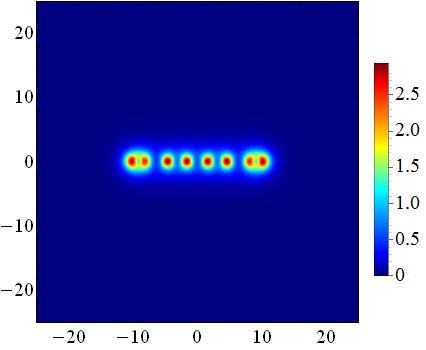} & \includegraphics[height=2.4cm,angle=00,bb=0 0 357 329]{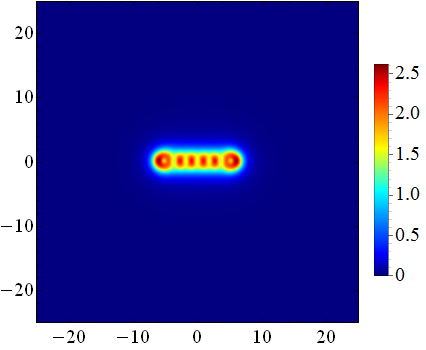} & \includegraphics[height=2.4cm,angle=00,bb=0 0 357 329]{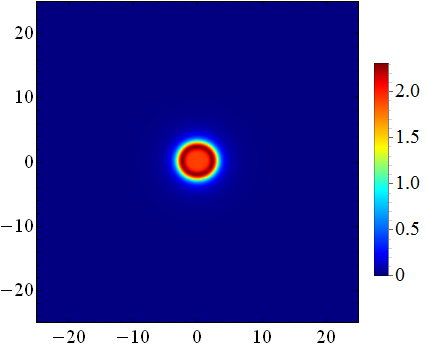} \\
        \hline
            9 & 0 & 0.4 & 0.6 & 1\\
& \includegraphics[height=2.4cm,angle=00,bb=0 0 357 329]{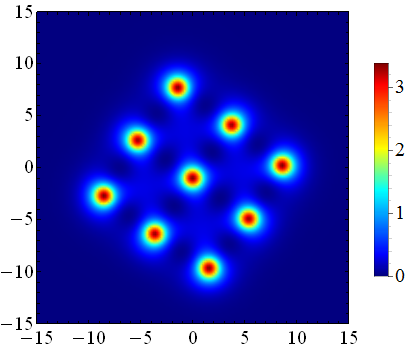} & \includegraphics[height=2.4cm,angle=00,bb=0 0 357 329]{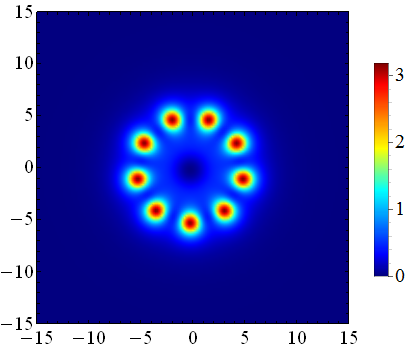} & \includegraphics[height=2.4cm,angle=00,bb=0 0 357 329]{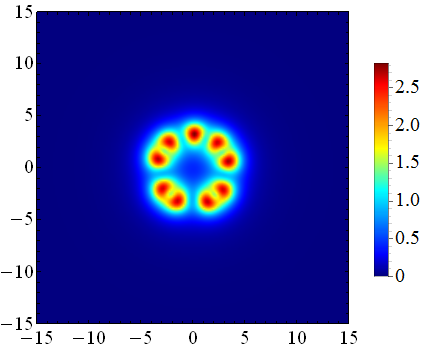} & \includegraphics[height=2.4cm,angle=00,bb=0 0 357 329]{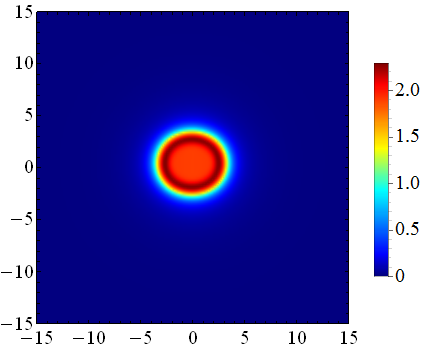} \\
        \hline
            10 & 0 & 0.4 & 0.6 & 1\\
& \includegraphics[height=2.4cm,angle=00,bb=0 0 357 329]{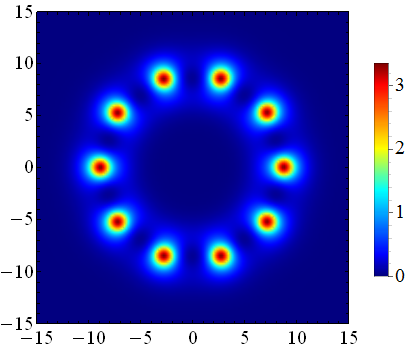} & \includegraphics[height=2.4cm,angle=00,bb=0 0 357 329]{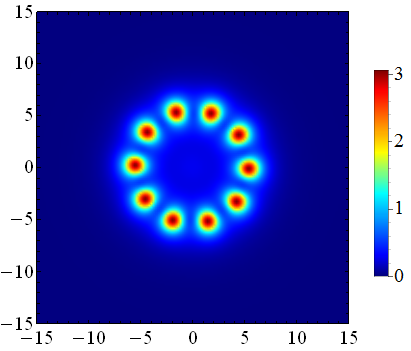} & \includegraphics[height=2.4cm,angle=00,bb=0 0 357 329]{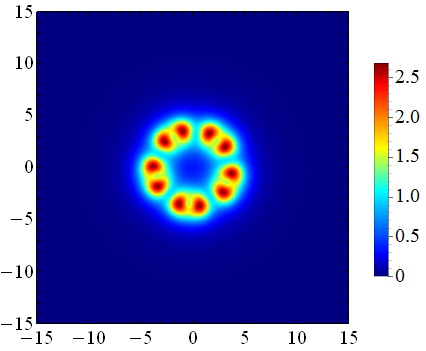} & \includegraphics[height=2.4cm,angle=00,bb=0 0 357 329]{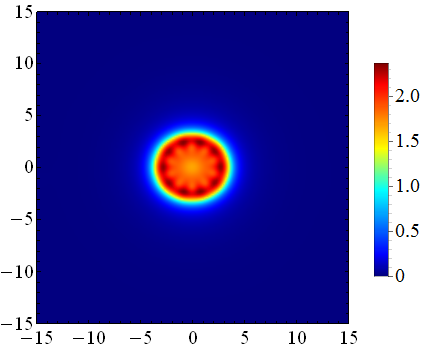} \\
        \hline
    \end{tabular}
    \caption{Energy density plots for $1\le Q\le 10$ gauged  planar Skyrmions in the model \re{lag} at some set of
    values of $g\in [0, 2]$ and $\mu^2=0.1$ and $\lambda=0.5$.}
 \lbfig{fig:3}
 \end{center}
\end{figure*}

Another check of the correctness of our results was
performed by finding  corresponding solutions of the Euler-Lagrange equations
which follow from the Lagrangian \re{lag} subject to a set of boundary conditions \re{bc}.
The relative errors of the solutions we found implementing the Newton-Raphson fourth order
finite difference method, are of order of $10^{-4}$ or smaller.

First, we recover the usual $Q=1$ solution. Its energy for $g=0$ is $E_1=20.25$
this is within $0.1\%$ of the previously known result \cite{Salmi:2014hsa}.
This similarity is another nice validation of our numerical algorithm.

Since there are variety of configurations of planar aloof Skyrmions,
in order to label different these multi-solitons we are using the following notation: $QD_n $
where $Q$ is the topological charge of the planar configuration or its building block,
and $D_n$ is the dihedral group of corresponding symmetry. For example, in the sector of degree 4
we have found the $4 D_4$ (which is the global minimum at $g=0$), $4 D_3$ and $4 D_2$ solutions.

Note, that among
solutions we constructed using the rational map input \re{input},
there are new configurations, which were not considered in \cite{Salmi:2014hsa} since they represent local minima.
In Fig.~\ref{fig:1} we exhibited the energy density plots of some of these new solutions, $3 D_3$, $4 D_3$ and $9 D_9$, respectively.

Constructing these configurations we vary relative orientation $\chi$
of the constituents in the internal space. Indeed, the
total potential of interaction between two neighboring baby Skyrmions is proportional to
$\cos \chi$ \cite{Bsk}, it is repulsive if the solitons
are in phase ($\chi=0$) and the most attractive channel corresponds to opposite orientations of
the solitons, $\chi = \pi$.

However, the interaction energy is still negative for the solitons which are
not completely out of phase. If the number of constituents is even, they always form pairs with
opposite relative orientation, however for a system of $N$ solitons, where $N$ is an odd number,
another possibility exist. The solitons may form a circular necklace where each of $N$ baby Skyrmions
is rotated by angle $(1-\frac{1}{N})\pi$ with respect to its neighbor.
Hence, in this case we considered composite solutions with $N$ soliton species, not only binary species
like in \cite{Salmi:2014hsa}. Thus, for odd $N$ the total angle of internal
rotation along the ring is $\pi (N-1)=2\pi k$, where $k$ is integer. Particular examples of this type,
$3D_3$ and  $9D_9$ solutions are displayed in Fig.~\ref{fig:1}. Clearly, configurations of that type
represent local minima.

Another interesting example of configuration, which also
was not discussed in \cite{Salmi:2014hsa}, is the "tristar"
configurations $4 D_3$ of degree $4$. Here the
relative phase of the 3 solitons with respect to the central constituent is $\pi$,
however, with respect to each other, they in phase, $\chi=0$.
Thus, the central component provides strong attraction to the outer solitons binding the configuration together.

As the gauge coupling increases from zero, the energy of the
gauged aloof Skyrmions decreases since the magnetic flux is
formed. Now each individual Skyrmion is coupled to a magnetic flux
and the electromagnetic interaction modifies the usual pattern of
scalar interaction of constituents. Effectively, the asymptotic field of the
gauged soliton represents a triplet of dipoles, two scalar dipole moments are
associated with asymptotic form of the scalar fields $\phi_1,\phi_2$ in the $x-y$ plane and
the third magnetic dipole is orthogonal to this plane \cite{Gladikowski:1995sc}.

\begin{figure}[hbt]
\lbfig{fig:2}
  \begin{center}
\begin{tabular}{cc}
\includegraphics[width=5.2cm, angle =-90]{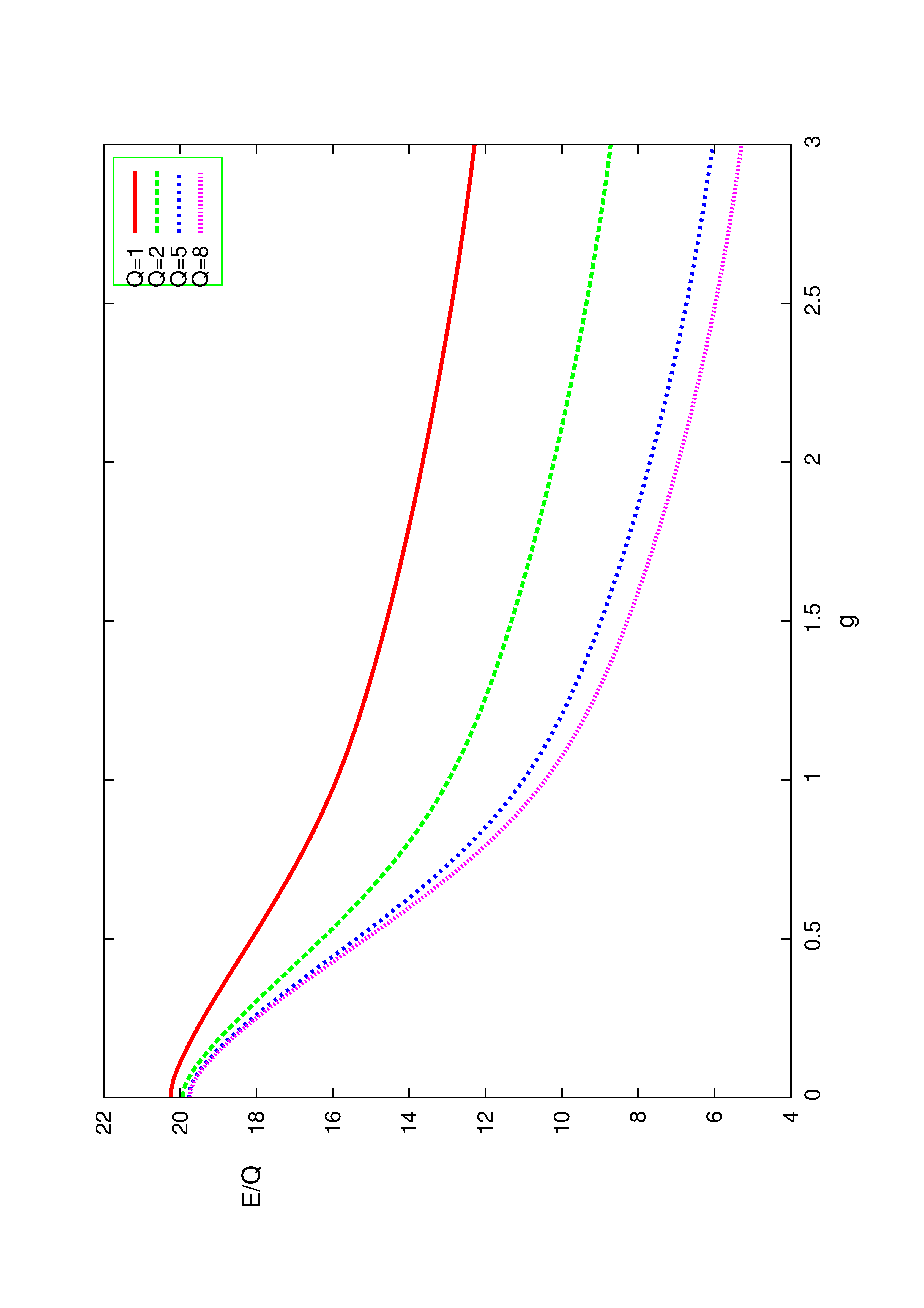} &
\includegraphics[width=5.2cm, angle =-90]{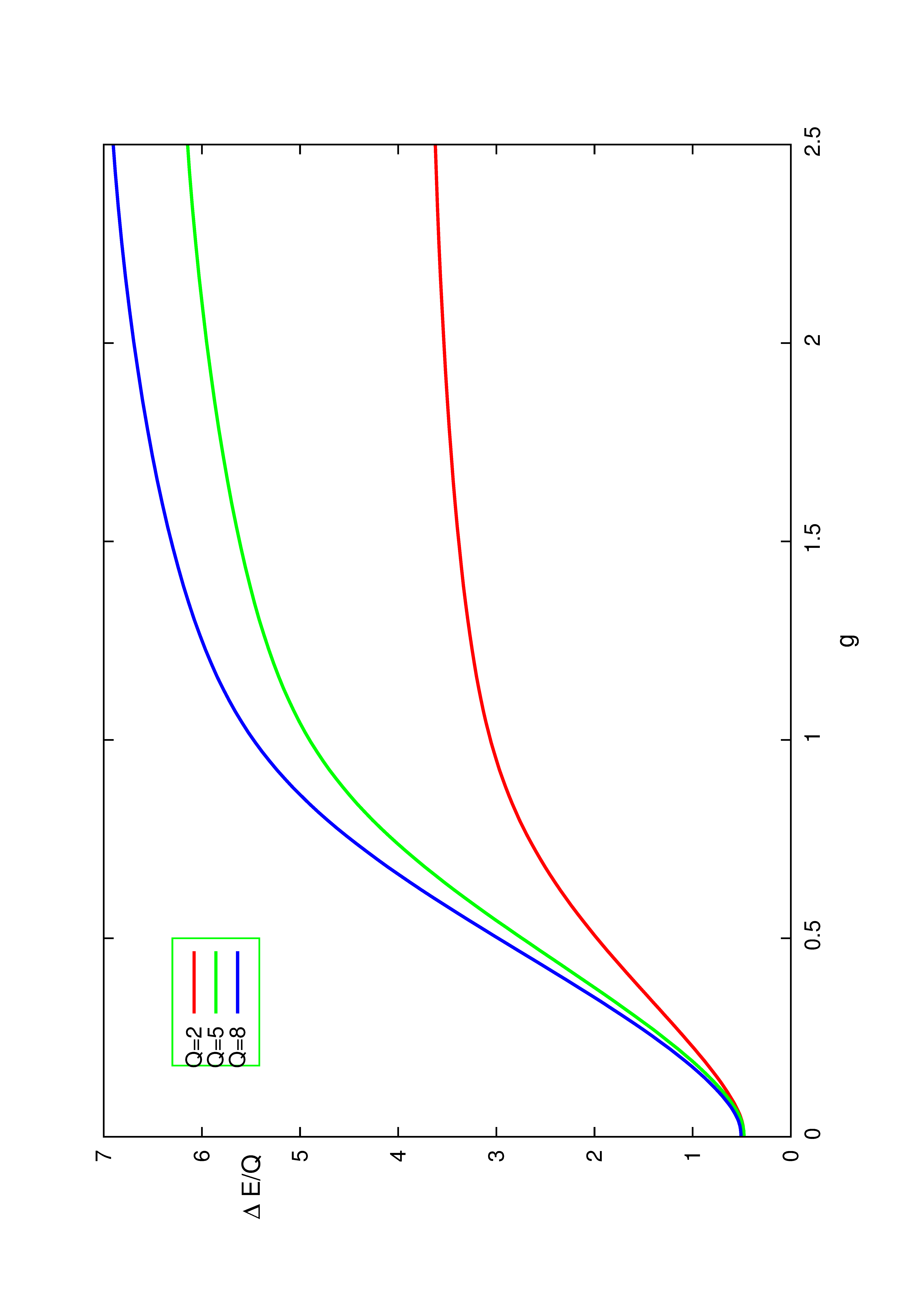}\\
\end{tabular}
\end{center}
\caption{\small The normalized energy per charge
$E$ of the $Q=1,2,5,8$ chains of gauged baby Skyrmions
(left plot), and the corresponding binding energy per charge (right plot)
as a function of the coupling constant $g$ at $\mu^2=0.1$ and $\lambda=0.5$. }
\end{figure}

Note that effectively, using the Maxwell
equation $\nabla \times \vec B = \vec j$, one can set the magnetic field
into correspondence with an effective circular electric current $j_\nu$. On the other hand the term $\sim g^2A^2 \phi^2$
in the total Hamiltonian of the system effectively contributes to the mass of the scalar field, thus the Yukawa
interaction between the gauged baby Skyrmions becomes stronger.

In Fig.~\ref{fig:3}  we exhibited the contour energy density plots of the
gauged baby Skyrmions in the model \re{lag} for $1\le Q\le 10$ at some set of values of the
gauge coupling $g \in [0.2]$.
First, we observe that for relatively small values of the
coupling constant $g$, two neighboring solitons
with opposite orientations form pairs in accordance with the a binary species model \cite{Salmi:2014hsa}.

Let us consider a few particular configurations.
In Fig.~\ref{fig:2}, left plot we have plotted the dependency of energy of the static gauged static baby Skyrmion of unit
charge and the chains $2D_2, 5D_2, 8D_2$ as functions of $g$. On the right plot we also
displayed the soliton's binding energy $\Delta E = Q E_1 - E$ as function of the gauge
coupling. Here $E_1$ is the energy of the one-soliton configuration at the corresponding value of the gauge coupling $g$
and we used the normalized units of energy per unit charge. From these plots it is clear that the energy of the configurations
decreases as the gauge coupling becomes stronger while the binding energy is increasing. Thus, the balance between the
repulsive and  attractive forces in the ungauged model with the potential \re{Salmi} is shifted
towards the attraction and, as coupling becomes strong enough, the rotational invariance of the multi-soliton
configuration is restored. Further increasing of the gauge coupling
makes the solitons width increasingly localized.

Note that as the coupling remains smaller than one, the electromagnetic energy
$E_{em}$  is increasing, however in the strong coupling limit its contribution begin
to decrease as $g$ continue to grow,  as seen in Fig.~\ref{fig:4}, left plot.

\begin{figure}[hbt]
\lbfig{fig:4}
\begin{center}
\begin{tabular}{cc}
\includegraphics[width=5.2cm, angle =-90]{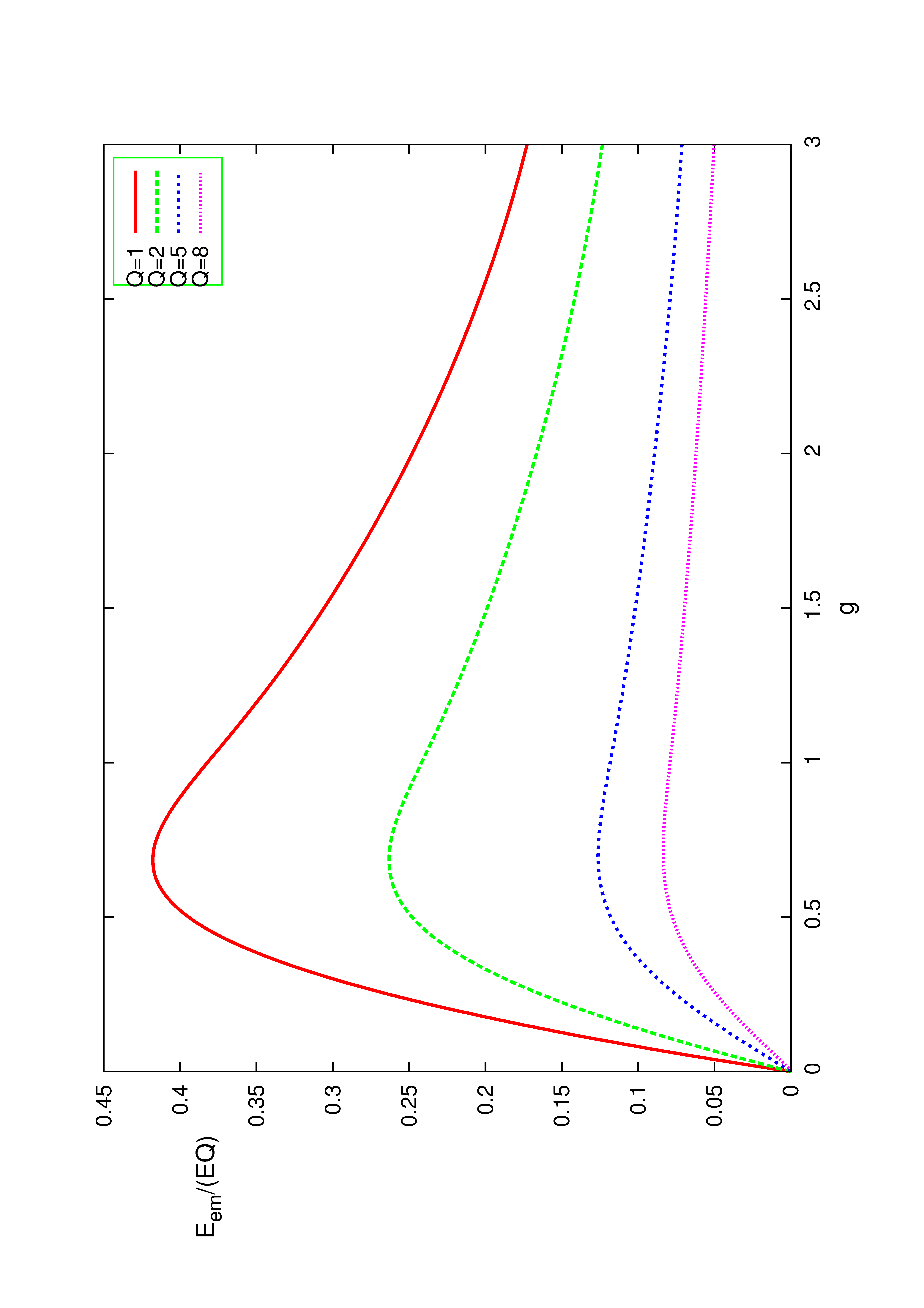} &
\includegraphics[width=5.2cm, angle =-90]{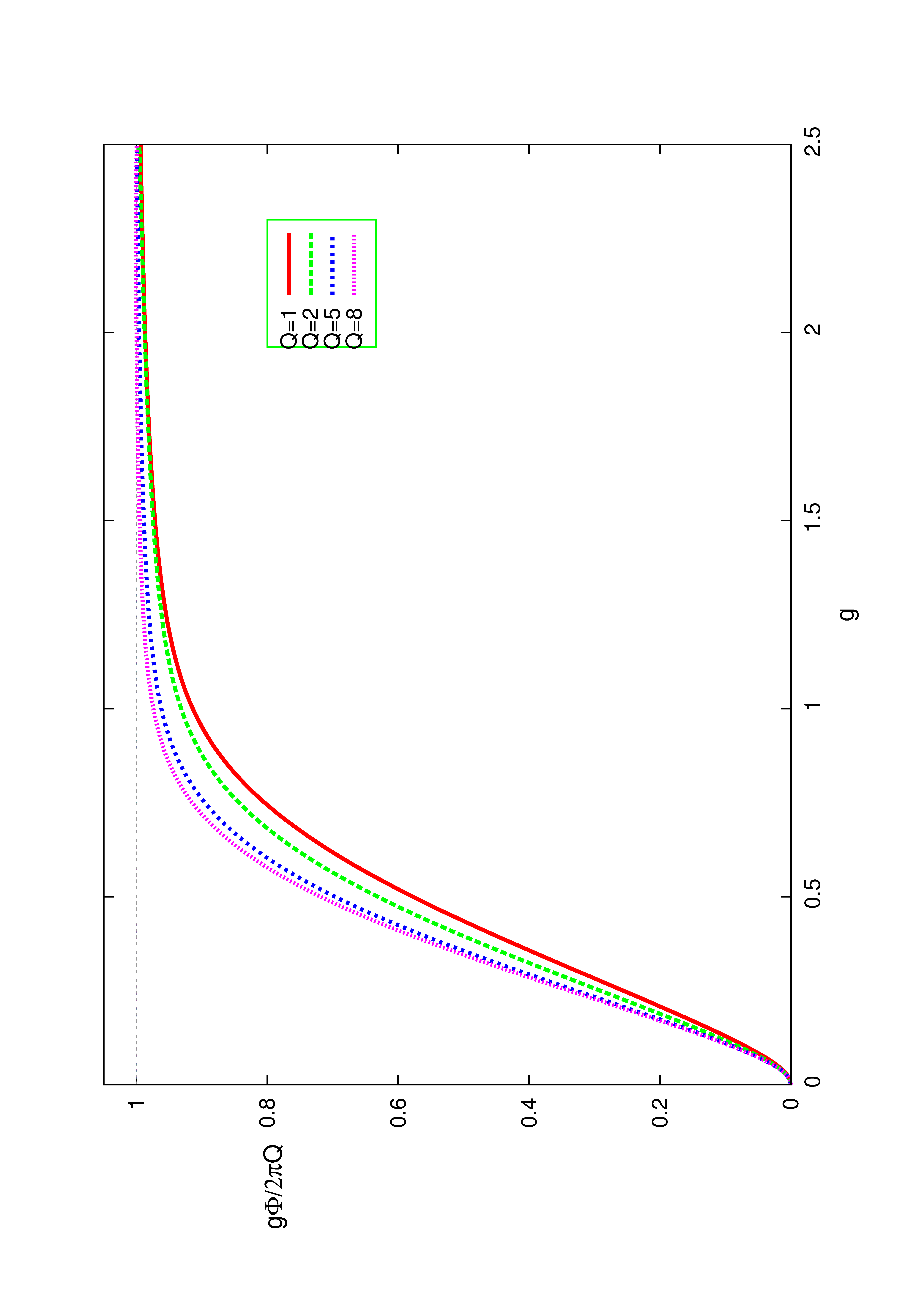}\\
\end{tabular}
\end{center}
\caption{\small The ratio per charge of the magnetic energy $E_{em}$ to the
total energy $E$ of the $Q=1,2,5,8$ chains of gauged baby Skyrmions
(left plot), and the magnetic flux through the $x-y$-plane  (right plot)
as a function of the coupling constant $g$ at $\mu^2=0.1$ and $\lambda=0.5$. }
\end{figure}

We can understand this effect if we note that the
conventional rescaling of the potential $A_\mu \to g A_\mu $
leads to $F_{\mu\nu}^2 \to \frac{1}{g^2}F_{\mu\nu}^2$. Thus, the very large gauge coupling effectively
removes the Maxwell term leaving the limiting configuration of gauged
planar Skyrmions coupled to a circular magnetic vortex of constant flux.
Apparently, in such a limit the strong coupling with a vortex yields an effective potential term which supplements the
potential \re{Salmi}.

Let us remind that the particular choice of the parameters of the aloof potential \re{Salmi} is related with the
condition of balance between a short-range repulsion and a long-range attraction, decreasing of the parameter
$\lambda$ increases the repulsive component. Coupling to magnetic field modifies the structure of interactions, as
the coupling $g$ increases, the attractive part of interaction becomes stronger. As a result, we can approach the limit
$\lambda \to 0$ providing the gauge coupling is non-vanishing. In such a limit, variation of the coupling $g$ allows us
to manipulate the attractive part of interaction between the Skyrmions.
Further, at some critical value of the gauge coupling there is a transition from the aloof structure of the multi-solitons
to the rotationally invariant configuration, for example in the sector of degree $Q=2$ it happens as $g=0.45$ and
$\lambda=0$.

Thus, the configurations carry total magnetic flux  $\Phi = \int d^2 x B$, which is in general, non-quantized.
The flux of the gauged baby Skyrmions is associated with the position of the solitons, it
is orthogonal to the $x-y$ plane \cite{Gladikowski:1995sc}.
In the usual model with rotationally invariant potential \re{double}, or in the strong coupling limit of the
gauge model \re{lag}, there is a single magnetic
flux through the center of the soliton, in the aloof system, where rotational invariance becomes violated,
each unit charge constituent of the multi-soliton configuration is coupled to a flux.

An interesting observation is that as the gauge coupling becomes stronger,
the magnetic flux of the degree $Q$ baby Skyrmions grows from 0 to $2\pi Q/g$, i.e. in the strong coupling regime
the magnetic flux is quantized thought there is no topological reasons for it \cite{Gladikowski:1995sc}.
Indeed, in Fig.~\ref{fig:4}, right plot, we display the results of our numerical calculations of the
integrated magnetic field of the gauged planar Skyrmion  through the $x-y$ plane. In the limit $g=0$ the magnetic
flux is vanishing, in the weak coupling regime $0\le g \lesssim \mu$ the fluxes are attached to the individual partons of the
multi-soliton configuration.

As gauge coupling increases further, the radius of the each vortex is getting smaller and the magnitude of the
magnetic field increases significantly. For $g>1$ the flux tends to be quantized in units of $2\pi/g$ and
the configuration becomes rotationally invariant. The energy density
distribution in the strong coupling limit is localized near the center of the
configuration approaching a singular string-like distribution in the limit $g\to \infty$ \cite{Gladikowski:1995sc}.

\section{Conclusion}
The main purpose of this work was to present new type of gauged solitons in the planar Skyrme-Maxwell theory.
In the model with aloof potential \re{Salmi} the individual solitons
are composed out of the constituents with unit topological charge.
Peculiar feature of the model is that, similar to the solutions of the Faddeev-Skyrme model,
a number of local energy minima of various types exists in each topological sector.
Coupling of the solitons to the gauge sector yields
additional attractive interaction which modifies the structure of configurations of given degree.
It has been shown that the rotational invariance of the multi-soliton configuration is restored in the
strong coupling regime.

Similar to the corresponding solutions in the gauged Skyrme model and Faddeev-Skyrme model
\cite{Shnir:2014mfa}, the planar configurations are topologically stable and in the weak coupling regime
they carry a non-quantized magnetic flux which is orthogonal to the $x-y$ plane and penetrates the Skyrmion.
In the strong coupling limit the total magnetic flux, associated with the partons,
becomes quantized in units of topological charge.

We confirm that the mass of the static configuration decreases when the electromagnetic coupling constant is increased,
thus a baby Skyrmion can lower its mass by interacting with the electromagnetic field.
Also when g is increasing, the binding energy of the solitons increases.

Finally, note that the planar Skyrmions appear as quasiparticles in various systems, in particular they are
natural objects at the description of the integer quantum Hall effect \cite{Girvin}. The vortices coupled to the
planar Skyrmion constituents may appear in
multicomponent superconductors \cite{Babaev}, thus
as avenues for further research, it would be interesting to extend the solutions in this work to the effective
condensed matter systems. Another interesting possibility is to consider the gauged baby Skyrme
model with aloof potential and Chern-Simons term \cite{Loginov} without restrictions of symmetry, this would allow us
to include the electric field into consideration.

\section*{Acknowledgments}
We thank Igor Bogolubsky, Derek Harland, Martin Speight and Andrzej Wereszczy\'nski
for useful discussions and valuable comments. This work is supported in part by the A. von
Humboldt Foundation in the framework of the Institutes linkage program
and by the JINR Heisenberg-Landau program (Y.S.).

\begin{small}

\end{small}
\end{document}